# Numerical implementation of flat FLRW models of cosmic expansion with Planck 2018 cosmological parameters

Bertua Marasca, J.C.[1]; Dávila Gutiérrez,H[1,2]; Martínez Osorio,O.A[1]; Mosquera Hadatty,J.I[1]; Videla Ricci,R.C[1].

[1]Master's Programme in Astrophysics and Astronomy, Universidad Internacional de La Rioja (UNIR), 26006 Logroño, Spain
[2]SKYCR, San José, Costa Rica

## ABSTRACT

We present an explicit numerical implementation of the Friedmann equations to describe the expansion of the Universe within spatially flat, homogeneous and isotropic Friedmann–Lemaître–Robertson–Walker (FLRW) models. We adopt up-to-date cosmological parameters from the Planck 2018 mission for the concordance ΛCDM model, including the present-day density fractions of radiation, matter and dark energy, as well as the value of the Hubble constant. Starting from the Friedmann acceleration equation together with the continuity equation for a perfect fluid with barotropic equation of state $p = w\rho$, we integrate the evolution of the scale factor $a(t)$ using an explicit Euler scheme implemented in Python. The code allows one to explore different constant values of the equation-of-state parameter $w$ associated with distinct energy components (radiation, non-relativistic matter, a cosmological constant and more exotic fluids), and to analyse how these values modify the history and fate of cosmic expansion. We present graphical results for the baseline ΛCDM scenario and for a family of models with $w \neq -1$, discussing their qualitative behaviour and their consistency with the standard cosmological paradigm. Finally, we highlight the potential of this implementation as a teaching tool in undergraduate and graduate cosmology courses, as well as its straightforward extension to more general models.

## 1. Introduction

The body of observational evidence accumulated since the pioneering work of Edwin Hubble has firmly established the picture of an expanding Universe. In particular, the approximately linear relation between galaxy distances and their radial velocities reported by Hubble (1929) is naturally interpreted, within general relativity, as the expansion of the space–time fabric itself (Carroll 2004). In this framework, the time evolution of the scale factor $a(t)$ lies at the heart of modern cosmology.

On this basis, the standard ΛCDM cosmological model has been developed. It describes a Universe that is currently dominated by cold dark matter and dark energy, with a smaller contribution from radiation. This model rests on several complementary observational pillars: the Hubble law, primordial nucleosynthesis of light elements, the large-scale structure of the Universe, and the properties of the cosmic microwave background (CMB) (Peacock 1999; Liddle 2015). High-precision observations from the Planck mission have yielded tight constraints on parameters such as the baryonic and dark-matter densities, the dark-energy density, the scalar spectral index, the spatial curvature and the age of the Universe (Planck Collaboration 2020).

In this context, numerical modelling of cosmic expansion plays a dual role. On the one hand, it provides a flexible way to explore the impact of different energy components and equations of state on the dynamics of a(t). On the other hand, it offers a direct bridge between the theoretical formalism and accessible computational implementations, which is particularly valuable in teaching environments (Nemiroff & Patla 2008; Baumann 2021).

The aim of this work is to present an explicit numerical implementation of the Friedmann equations for a flat FLRW Universe that incorporates cosmological parameters consistent with the Planck 2018 results. The model enables a unified treatment of the contributions from radiation, non-relativistic matter and dark energy with a constant equation of state $p = w\rho$, exploring both the reference ΛCDM case ($w = -1$) and alternative values of w that lead to qualitatively different expansion histories. The code is implemented in Python and has been designed to be easily reproducible and modifiable.

## 2. Theoretical framework

### 2.1. Cosmological principle and FLRW metric

On sufficiently large scales, observations indicate that the Universe is approximately homogeneous and isotropic. The cosmological principle states that, on these scales, the Universe is statistically the same at every point and in every direction (Peacock 1999; Liddle 2015). Under these assumptions the space–time metric is restricted to the Friedmann–Lemaître–Robertson–Walker (FLRW) family (Carroll 2004; Baumann 2021).

In comoving coordinates (t, r, θ, φ) and adopting units in which $c = 1$, the FLRW line element can be written as in equation (1):

FLRW line element.

$$(ds)^2 = (dt)^2 - a(t)^2\left[\frac{(dr)^2}{1-kr^2} + r^2(d\theta)^2 + r^2\sin^2\quad\theta,(d\phi)^2\right] \quad (1)$$

Here, a(t) is the scale factor and k is the spatial curvature parameter, which can take the values $k = 0$ (flat space), $k = +1$ (closed space) or $k = -1$ (open space). The time dependence of the geometry is entirely encoded in a(t), which measures how comoving distances between galaxies change with time. It is customary to normalize the present-day value to $a(t_0) = 1$ for convenience.

### 2.2. Einstein equations and Friedmann equations

The dynamics of space–time in general relativity are governed by Einstein's field equations (Carroll 2004):

Einstein field equations.

$$R_{\mu\nu} - \frac{1}{2} g_{\mu\nu} R + \Lambda g_{\mu\nu} = 8\pi G, T_{\mu\nu} \qquad (2)$$

In cosmology it is common to model the matter content of the Universe as a perfect fluid, characterised by an energy density $\rho$, an isotropic pressure $p$, and a four–velocity $u^\mu$ (Carroll 2004):

Perfect–fluid energy–momentum tensor.

$$T_{\mu\nu} = (\rho + p) u_\mu u_\nu - p, g_{\mu\nu} \qquad (3)$$

The relation between pressure and energy density is specified by a barotropic equation of state,

Equation of state.

$$p = \omega, \rho \quad (4)$$

Different values of $w$ correspond to different physical components: non-relativistic matter ($w = 0$), radiation ($w = 1/3$), vacuum energy or a cosmological constant ($w = -1$), as well as possible exotic forms of energy when $w$ takes other values (Peacock 1999; Liddle 2015).

By inserting the FLRW metric (equation 1) and the perfect-fluid energy–momentum tensor (equation 3) into Einstein's equations (equation 2), one obtains the system of differential equations known as the Friedmann equations (Carroll 2004; Romeu 2014). The Hubble parameter is defined as

Hubble parameter.

$$H(t) = \frac{\dot{a}(t)}{a(t)} \quad (5)$$

where an overdot denotes a derivative with respect to cosmic time.

The first Friedmann equation can be written as

First Friedmann equation.

$$\left(\frac{\dot{a}}{a}\right)^2 = \frac{8\pi G}{3}, \rho - \frac{k}{a^2} + \frac{\Lambda}{3} \qquad (6)$$

The second Friedmann equation takes the form

Second Friedmann equation.

$$\frac{\ddot{a}}{a} = -\frac{4\pi G}{3}(\rho + 3p) + \frac{\Lambda}{3} \qquad (7)$$

On the other hand, local conservation of energy–momentum, $\nabla_\mu T^{\mu\nu} = 0$, leads to the cosmological continuity equation (Carroll 2004):

Continuity equation.

$$\dot{\rho} + 3\frac{\dot{a}}{a}(\rho + p) = 0 \quad (8)$$

For a fluid with a barotropic equation of state $p = w\rho$ and constant w, the solution of the continuity equation (8) is (Carroll 2004; Baumann 2021):

General density scaling.

$$\rho(a) = \rho_0, a^{-3(1+\omega)} \quad (9)$$

This expression shows how the energy density of each component decays as the Universe expands: non-relativistic matter scales as $a^{-3}$, radiation as $a^{-4}$, while a cosmological constant ($w = -1$) remains constant in time.

**2.3. Energy components and Planck 2018 parameters**

In the standard ΛCDM model, the total energy density of the Universe is decomposed into radiation ($\Omega_r$), matter ($\Omega_m$, including baryonic matter and cold dark matter), and dark energy ($\Omega_\Lambda$), with a possible additional term associated with spatial curvature ($\Omega_k$):

Sum of density parameters.

$$\Omega_r + \Omega_m + \Omega_\Lambda + \Omega_k = 1 \qquad (10)$$

Observations from the Planck 2018 mission indicate that the Universe is very nearly spatially flat ($\Omega_k \simeq 0$), with representative values (Planck Collaboration 2020):
$\Omega_m \simeq 0.315$,
$\Omega_r \simeq 9.24 \times 10^{-5}$,
$\Omega_\Lambda \simeq 0.685$,
$H_0 \simeq 67.4$ km $s^{-1}$ $Mpc^{-1}$.

In terms of the Hubble parameter H(t) defined in equation (5), the first Friedmann equation (6) for a flat Universe ($k = 0$) can be rewritten as (Peacock 1999; Liddle 2015):

H(a) in a flat Universe.

$$H(a)^2 = H_0^2\left[\Omega_r a^{-4} + \Omega_m a^{-3} + \Omega_{de} a^{-3(1+\omega_{de})}\right] \qquad (11)$$

where $\Omega_{de}$ denotes the density fraction associated with the dark-energy component and $w_{de}$ its effective equation-of-state parameter. For a pure cosmological constant one has $w_{de} = -1$.

The present-day critical density is defined as (Carroll 2004):

Critical density.

$$\rho_c = \frac{3H_0^2}{8\pi G} \qquad (12)$$

The present-day densities of each component can then be written as

Present-day matter density.

$$\rho_{m,0} = \Omega_m, \rho_c \quad (13)$$

The present–day radiation density is

$$\rho_{r,0} = \Omega_r, \rho_c \quad (14)$$

and the present–day dark–energy density is

$$\rho_{de,0} = \Omega_{de}, \rho_c \quad (15)$$

Applying the general result (9) to each component, we obtain their dependence on the scale factor:

Radiation

$$\rho_r(a) = \rho_{r,0}, a^{-4} \qquad (16)$$

Non-relativistic matter

$$\rho_m(a) = \rho_{m,0}, a^{-3} \qquad (17)$$

Dark energy with constant w

$$\rho_{de}(a) = \rho_{de,0}, a^{-3(1+\omega)} \quad (18)$$

The sum of all contributions gives the total energy density,

3. $\rho(a) = \rho_r(a) + \rho_m(a) + \rho_{de}(a) \qquad (19)$

## 4. Numerical model and Python implementation

### 3.1. Model set-up

Using the expressions above, we aim to study the evolution of the scale factor a(t) for a homogeneous, isotropic and spatially flat Universe (k = 0) composed of radiation, non-relativistic matter and dark energy with an effective equation-of-state parameter w. The total energy density is given by equation (19), with the specific scalings (16)–(18).

We start from the second Friedmann equation (7). For each component we have p_i = w_i ρ_i, with w_r = 1/3 for radiation, w_m = 0 for non-relativistic matter, and w for the dark-energy component. The combination (ρ + 3p) can then be written as (Carroll 2004; Nemiroff & Patla 2008)

$$\rho + 3p = 2\rho_r(a) + \rho_m(a) + (1 + 3\omega)\rho_{de}(a) \qquad (20)$$

Substituting equation (20) into equation (7) and absorbing the cosmological constant $\Lambda$ into the dark-energy component, we obtain the acceleration equation used in our model,

Acceleration equation (divided by a)

$$\frac{\ddot{a}}{a} = -\frac{4\pi G}{3}\left[2\rho_r(a) + \rho_m(a) + (1 + 3\omega)\rho_{de}(a)\right] \qquad (21)$$

or, equivalently, in explicit form for a(t),

Explicit acceleration equation

$$\ddot{a}(t) = -\frac{4\pi G}{3}\left[2\rho_r(a) + \rho_m(a) + (1 + 3\omega)\rho_{de}(a)\right]a(t) \qquad (22)$$

This equation (22), together with the expressions (16)–(18), forms the basis of our numerical model.

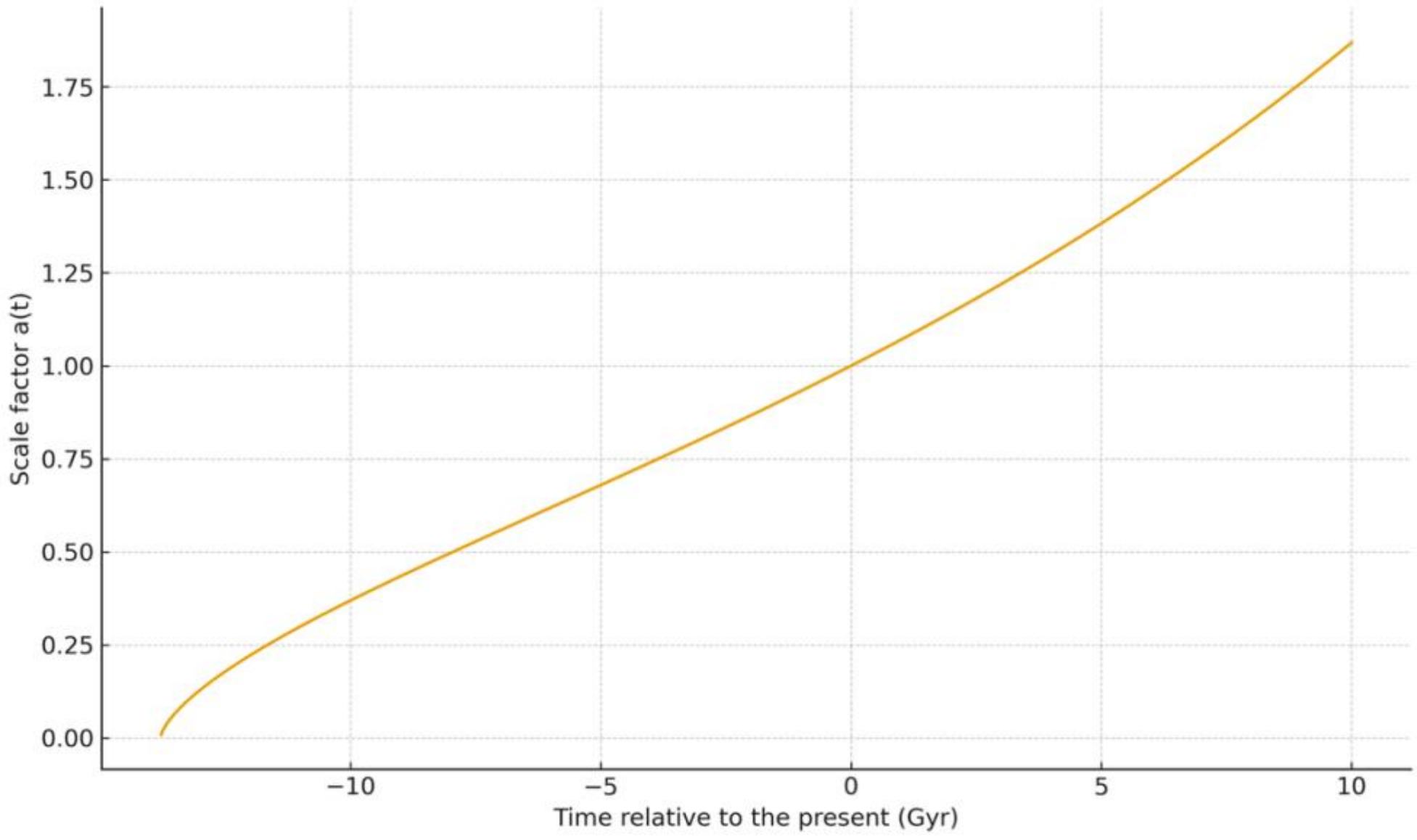

Figure 1. Evolution of the scale factor a(t) in a spatially flat ΛCDM universe, obtained by numerically integrating the Friedmann acceleration equation with Planck 2018 cosmological parameters ($\Omega m = 0.315$, $\Omega r = 9.24 \times 10^{-5}$, $\Omega\Lambda = 0.685$, $H_0 = 67.4$ km $s^{-1}$ $Mpc^{-1}$). Time is measured relative to the present epoch (t = 0) in gigayears. The model assumes that dark energy is described by a cosmological constant with equation-of-state parameter $w = -1$. The curve illustrates the transition from an early decelerated expansion, dominated by radiation and then matter, to the late-time accelerated phase driven by dark energy.

### 3.2. Reformulation as a first–order system

To solve the second–order equation (22) numerically, it is convenient to recast the problem as a system of first–order equations (Baumann 2021; Nemiroff & Patla 2008). We introduce the auxiliary variable

Expansion–velocity variable

$$v(t) = \dot{a}(t) \quad (23)$$

so that the dynamical system becomes

Equation for $a(t)$

$$\frac{da}{dt} = v(t) \quad (24)$$

Equation for $v(t)$

$$\frac{dv}{dt} = -\frac{4\pi G}{3}[2\rho_r(a) + \rho_m(a) + (1 + 3\omega)\rho_{de}(a)]a(t) \quad (25)$$

The initial conditions are specified at the present cosmic time $t_0$, which we take to be $t_0 = 0$. It is customary to impose (Carroll 2004; Planck Collaboration 2020):

$a(0) = 1$
$v(0) = H_0$

thereby normalising the model to the present-day value of the Hubble constant.

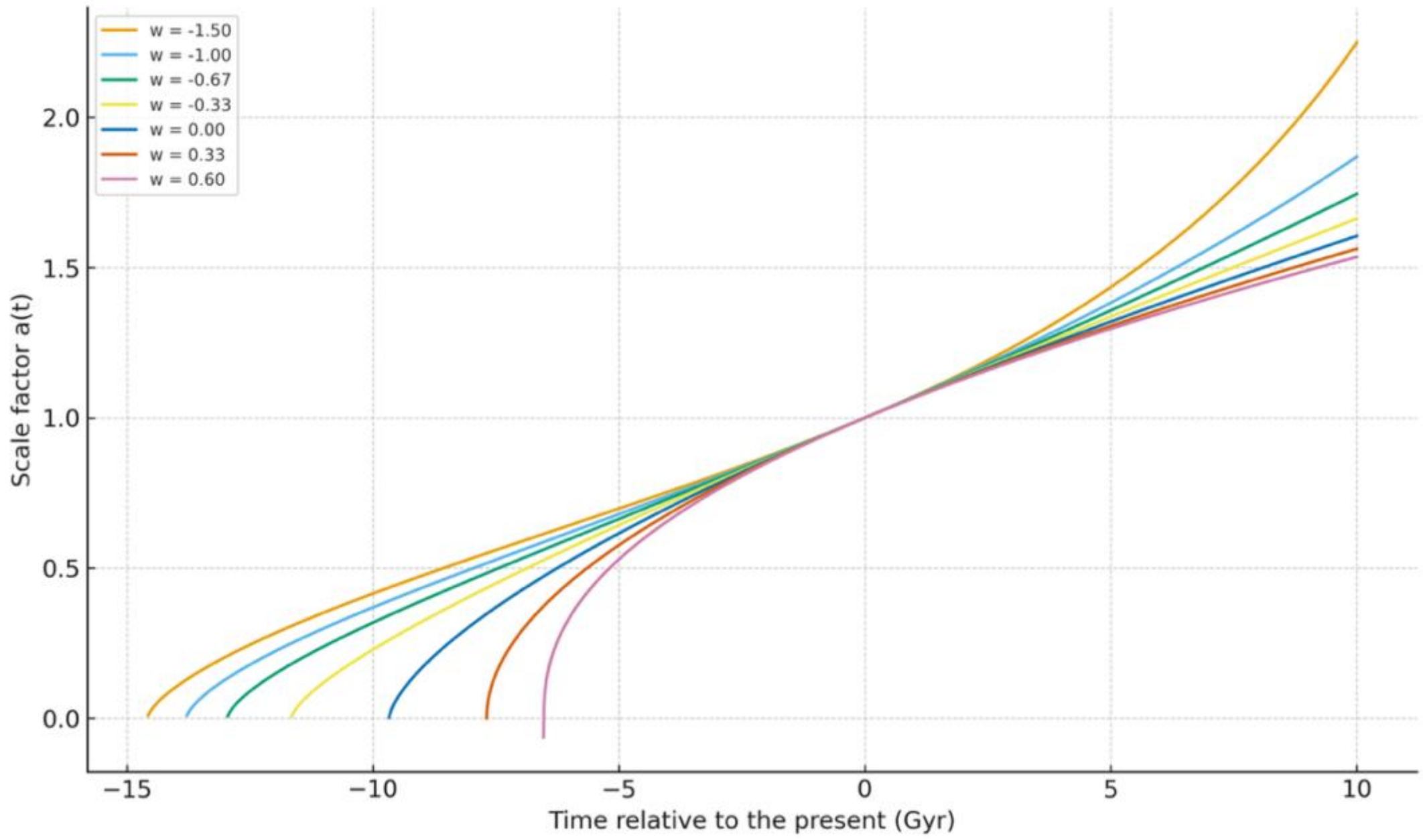

Figure 2. Comparison of the scale factor a(t) for different constant values of the dark energy equation-of-state parameter w, keeping the present-day density parameters fixed to the Planck 2018 values ($\Omega m = 0.315$, $\Omega r = 9.24 \times 10^{-5}$, $\Omega\Lambda = 0.685$). All models are normalized such that $a(t_0) = 1$ at the present time $t_0 = 0$. Curves with $w < -1$ (phantom energy) exhibit a more rapidly accelerating expansion, potentially leading to a future “Big Rip” scenario, whereas models with $w = -1$ reproduce the standard ΛCDM behaviour. Intermediate cases such as $w = -2/3$ (domain walls) and $w = -1/3$ (cosmic strings) produce milder acceleration, while $w = 0$ (pressureless matter) and $w = 1/3$ (radiation) correspond to non-accelerating power-law expansions. The ultralight case $w = 0.6$ is included as a hypothetical example with a very stiff equation of state, mainly for illustration.

## 3.3. Physical parameters and units

In the numerical implementation we adopt representative Planck 2018 values (Planck Collaboration 2020):

$$
\begin{aligned}
&H_0 = 67.4\ \mathrm{km\ s^{-1}\ Mpc^{-1}},\\
&\Omega_m = 0.315,\\
&\Omega_r = 9.24 \times 10^{-5},\\
&\Omega_{de} = 0.685.
\end{aligned}
$$

Using these values, the critical density is computed from equation (12), and the present-day densities of each component are then obtained via equations (13)–(15). In the code, $H_0$is converted to time units consistent with the integration scale (years or gigayears, Gyr), and a time step $\Delta t$is chosen small enough to provide a stable integration and a good resolution of the evolution of $a(t)$. In our numerical experiments we employ time steps of order $10^7$years to explore intervals up to $10^{10}$years both into the future and into the past.

## 3.3. Physical parameters and units

In the numerical implementation we adopt representative Planck 2018 values (Planck Collaboration 2020):

$$
\begin{aligned}
&H_0 = 67.4\ \mathrm{km\ s^{-1}\ Mpc^{-1}},\\
&\Omega_m = 0.315,\\
&\Omega_r = 9.24 \times 10^{-5},\\
&\Omega_{de} = 0.685.
\end{aligned}
$$

Using these values, the critical density is computed from equation (12), and the present-day densities of each component are then obtained via equations (13)–(15), i.e. $\rho_{i,0} = \Omega_i \rho_c$. In the code, $H_0$is first converted to SI units (s $^{-1}$) and then to yr $^{-1}$, so that cosmic time can be handled conveniently in years and gigayears (Gyr).

We adopt a fixed time step $\Delta t = 10^7$yr and integrate the equations both backwards and forwards in time. Towards the future, we explore intervals up to $10^{10}$yr from the present epoch, whereas towards the past the integration is stopped once the scale factor reaches $a \simeq 0.01$. This avoids the immediate vicinity of the initial singularity, where the classical FLRW description ceases to be applicable.

Keeping the total time span fixed, we have checked that reducing the time step does not appreciably change the morphology of the resulting $a(t)$curves. This indicates that the explicit Euler scheme provides a qualitatively stable approximation for the purposes of this work. Nevertheless, higher–order methods, such as fourth–order Runge–Kutta, would be preferable in applications requiring greater quantitative accuracy (Nemiroff & Patla 2008; Baumann 2021).

## 3.4. Numerical scheme and code structure

The Python code implements an explicit Euler integration scheme. Although higher–order methods such as Runge–Kutta offer improved accuracy, the Euler method is well suited as a first approximation and has considerable pedagogical value (Nemiroff & Patla 2008). The overall structure of the algorithm can be summarised in the following steps:

1. Define the physical constants $(G, H_0)$and the cosmological parameters $(\Omega_m, \Omega_r, \Omega_{de})$according to Planck 2018.
2. Compute the critical density $\rho_c$from equation (12) and the present-day densities of each component using equations (13)–(15).
3. Set the initial conditions $a(0) = 1$and $v(0) = H_0$.
4. Choose the time step $\Delta t$and the total integration interval into the past and into the future.
5. For each time step:
   (a) Evaluate $\rho_r(a)$, $\rho_m(a)$and $\rho_{de}(a)$from equations (16)–(18).
   (b) Compute $dv/dt$using equation (25).
   (c) Update the "velocity" $v$:
   $$v(t + \Delta t) = v(t) + \left(\frac{dv}{dt}\right)\Delta t.$$
   (d) Update the scale factor:
   $$a(t + \Delta t) = a(t) + v(t)\Delta t.$$
   (e) Store the values of $t$and $a(t)$for later plotting.

The full source code can be provided in an appendix or in a public repository, while the main text summarises the method as outlined above.

## 4. Results and discussion

### 4.1. Baseline case: ΛCDM Universe

As a first case study we consider the standard scenario in which dark energy is modelled as a cosmological constant, with $w = -1$(Carroll 2004; Liddle 2015; Planck Collaboration 2020). In this case the dark–energy density remains constant in time, whereas the matter and radiation densities decay as $a^{-3}$and $a^{-4}$, respectively, in agreement with equations (16) and (17).

Our numerical simulations show that at early times (small values of $a$) the dynamics are dominated by radiation, followed by a matter–dominated phase and, at late times, by dark energy. This behaviour is consistent with the approximate analytic solutions of the Friedmann equation (Carroll 2004; Liddle 2015):

Radiation–dominated Universe

$$a(t) \propto t^{1/2} \quad (26)$$

Matter–dominated Universe

$$a(t) \propto t^{2/3} \quad (27)$$

Cosmological–constant–dominated Universe

$$a(t) \propto e^{Ht} \quad (28)$$

In the numerical model, integrating the system (24)–(25) with $w = -1$and the Planck 2018 parameters, the resulting curve $a(t)$qualitatively reproduces this sequence: an initial phase of relatively slow expansion (radiation–dominated regime), followed by a regime approximately proportional to $t^{2/3}$(matter domination), and finally a phase of accelerated expansion with an almost exponential behaviour (dark–energy domination).

### 4.2. Exploring different values of $w$

The value of $w$in the dark–energy component plays a crucial role in the evolution of the Universe and its ultimate fate (Peacock 1999; Carroll 2004; Baumann 2021). By modifying $w$in equations (18), (20) and (25), the model allows us to explore numerically a variety of scenarios:

• Phantom energy ($w < -1$): the pressure is so negative that the energy density increases as the Universe expands. The acceleration equation (21) shows that the term $(1+3w)$can become strongly negative, leading to extremely rapid accelerated expansion. In this regime, "Big Rip"–type scenarios may arise, in which the scale factor diverges in a finite cosmic time and even bound structures could eventually be disrupted.

• Vacuum energy or cosmological constant ($w = -1$): this corresponds to the ΛCDM case. The dark–energy density remains constant, and the simulations display accelerated but moderate expansion, with $a(t)$tending towards exponential growth, as in equation (28).

• Domain walls ($w = -2/3$) and cosmic strings ($w = -1/3$): these values can be associated with two–dimensional and one–dimensional topological defect configurations, respectively (Peacock 1999). In our model, the corresponding terms in equation (21) produce expansion histories with intermediate degrees of acceleration, lying between the matter–dominated and cosmological–constant cases.

• Non-relativistic matter (w = 0) and radiation (w = 1/3): when we consider Universes dominated by these components (i.e. setting ρ_de = 0 in equation 19), the numerical solutions reproduce the power–law behaviours (26) and (27).

• Hypothetical components with w > 1/3: these represent fluids with a pressure even larger than that of radiation. From a theoretical standpoint such models are constrained by causality considerations (the sound speed should not exceed the speed of light), but the numerical scheme allows their behaviour to be explored in a controlled way (Baumann 2021).

The resulting curves for a(t) in each of these cases exhibit qualitatively different trends, ranging from non–accelerating expansions (matter, radiation) to exponential or even super–exponential growth (cosmological constant, phantom energy). The visual comparison of these curves helps students to understand how the parameter w controls the expansion rate and the ultimate fate of the Universe.

### 4.3. Qualitative comparison with the ΛCDM model

The simulation based on w = −1 and the Planck 2018 parameters shows clear qualitative agreement with the concordance ΛCDM model (Planck Collaboration 2020). The transition between radiation–, matter– and dark–energy–dominated regimes emerges naturally when integrating equations (24)–(25) with the energy densities (16)–(18) and the combination (20).

Moreover, exploring values of w different from −1 illustrates how deviations from a pure cosmological constant would modify cosmic expansion. For $w > -1$ the expansion is either accelerated but weaker than in the ΛCDM case, or becomes non–accelerating for sufficiently large values (e.g. $w = 0$ or 1/3). For $w < -1$ the acceleration becomes more extreme and may lead to cosmological instabilities such as Big–Rip scenarios (Carroll 2004; Baumann 2021).

Although current observations favour values close to $w = -1$ (Planck Collaboration 2020), the numerical model presented here provides a flexible framework in which to experiment with alternative hypotheses and to gain a better understanding of how observational data constrain the physics of dark energy.

## 5. Conclusions

In this work we have developed a numerical model of the expansion of the Universe within the FLRW framework, incorporating cosmological parameters consistent with the Planck 2018 results (Planck Collaboration 2020). Starting from the Friedmann equations (6) and (7) and the continuity equation (8), we have implemented in Python the evolution of the scale factor a(t) for a flat Universe composed of radiation, non–relativistic matter and dark energy with equation of state $p = w\rho$.

The main results of this work can be summarised as follows:

1. The numerical implementation of the system (24)–(25), together with the density scalings (16)–(18), qualitatively reproduces the expected behaviour of the standard ΛCDM model: an initial radiation–dominated phase, followed by a matter–dominated regime and, finally, a late epoch of accelerated expansion driven by dark energy with $w \simeq -1$(Carroll 2004; Liddle 2015; Planck Collaboration 2020).
2. Exploring different values of the equation–of–state parameter $w$shows how small deviations around $w = -1$modify the strength of the accelerated expansion, whereas more exotic values ($w < -1$ or $w > 1/3$) lead to radically different cosmological scenarios, such as possible Big–Rip endings in the case of phantom energy (Carroll 2004; Baumann 2021).
3. From a teaching perspective, the model is particularly useful. The Python code has moderate complexity and can be used in cosmology courses to help students connect physical parameters (energy densities, $w$, $H_0$) directly with the global behaviour of the Universe. The possibility of easily changing these parameters and observing their impact on the $a(t)$curves facilitates the understanding of abstract concepts such as dark energy and the fate of the Universe (Nemiroff & Patla 2008; Baumann 2021).
4. The approach presented here can be extended in several directions: including spatial curvature ($\Omega_k \neq 0$) in equation (11), considering models with dynamical dark energy $w(t)$in equation (9), and adopting more advanced numerical schemes (e.g. fourth–order Runge–Kutta) to improve the accuracy and stability of the integration (Baumann 2021).

Such implementations can directly complement the usual analytical treatments in cosmology courses, providing students with a concrete tool to explore how cosmic expansion responds to the physical parameters of the model.

Appendices

3.5. Simulation code (Appendix 6)

The numerical integration of the cosmic expansion is implemented in a Python script, whose full listing is given in Appendix 6. The code was run in a Jupyter Notebook using the `%matplotlib inline` directive, and makes use of the standard libraries `math`, `numpy`, and `matplotlib`.

First, the physical and cosmological constants are defined according to the Planck 2018 results: the present-day Hubble constant $H_0$, initially expressed in km $\mathrm{s}^{-1}$ $\mathrm{Mpc}^{-1}$and converted to units of $\mathrm{yr}^{-1}$; the gravitational constant $G$in units of $\mathrm{m}^3$ $\mathrm{kg}^{-1}$ $\mathrm{yr}^{-2}$; and the density parameters $\Omega_m$, $\Omega_r$, and $\Omega_\Lambda$. The critical density $\rho_c$is then computed from equation (12), and the present-day densities of matter, radiation, and dark energy are obtained from equations (13)–(15) and stored in the variables `densidadMateria`, `densidadRadiacion`, and `densidadEnergia`, respectively.

The equation-of-state parameter of the dark-energy component, $w$, is initially set to $w = -1$, corresponding to a cosmological constant, but it can be easily modified to explore different scenarios. The time step is specified as `incrementoTemporal = 1.0e7` years, which allows the acceleration equation (22) to be integrated using an explicit Euler scheme.

The code integrates separately towards the past ($t < 0$) and towards the future ($t > 0$). For $t < 0$, it starts from the initial conditions $a(0) = 1$and $\dot{a}(0) = H_0$, represented in the code by `factorEscalaActual = 1.0` and `hubbleActual = constanteHubble`. At each iteration the acceleration $\mathrm{d}^2 a/\mathrm{d}t^2$is evaluated using equation (22), directly employing the density scalings and their dependence on $a$. The variables `hubbleActual` (which in the code plays the role of $\dot{a}$) and `factorEscalaActual` are then updated with a negative time step until the scale factor reaches $a \simeq 0.01$. The pairs $(t, a(t))$, with time expressed in gigayears, are stored in the list `resultadoPasado`.

An analogous procedure is followed for $t > 0$, now with a positive time step, up to a maximum time of 10 Gyr into the future. The results are stored in the list `resultadoFuturo`. Finally, the data from both lists are combined into a single array containing the complete evolution of $a(t)$around the present epoch, and the scale factor is plotted as a function of time, with the present set at $t = 0$.

This code directly implements the system of equations (24)–(25), making use of the acceleration equation (22) and the scale-factor dependences given by (16)–(18). It constitutes the computational core of the simulations presented in Section 4.

Appendix 6. Python code (clean version consistent with the text)

The listing below contains the exact Python code used in our Jupyter Notebook simulations.

## Appendix 6. Python code for the FLRW expansion simulations

```
# Simulación de la expansión del universo con parámetros de Planck 2018
# Universo plano con radiación, materia y energía oscura.
# Integración numérica de a(t) mediante esquema de Euler explícito.

%matplotlib inline

import math
```

```
import numpy as np
import matplotlib.pyplot as plt

# ----------------------------------------------------------------------
# Constantes (Planck 2018) y conversiones
# ----------------------------------------------------------------------

# Segundos por año
seg_por_yr = 365.25 * 24 * 3600  # [s/yr]

# Constante de Hubble actual: H0 = 67.4 km/s/Mpc -> en 1/s
H0_si = 67.4e3 / 3.08567758e22   # [s^-1]

# Constante de Hubble en unidades de 1/año
constanteHubble = H0_si * seg_por_yr  # [yr^-1]

# G en SI [m^3 kg^-1 s^-2] -> [m^3 kg^-1 yr^-2]
G_si = 6.67408e-11
constanteGravedad = G_si * seg_por_yr**2  # [m^3 kg^-1 yr^-2]

# Densidad crítica: rho_c = 3 H0^2 / (8 π G)
rho_crit = 3.0 * H0_si**2 / (8.0 * math.pi * G_si)  # [kg m^-3]

# Parámetros de densidad (Planck 2018)
Omega_M = 0.315     # materia (bariónica + fría)
Omega_r = 9.24e-5   # radiación
Omega_L = 0.685     # energía oscura (tipo constante cosmológica)

# Densidades actuales de cada componente
densidadMateria   = Omega_M * rho_crit   # [kg m^-3]
densidadRadiacion = Omega_r * rho_crit   # [kg m^-3]
densidadEnergia   = Omega_L * rho_crit   # [kg m^-3]

# Parámetro de ecuación de estado de la energía oscura (p = ω ρ)
omega = -1.0  # se puede variar para explorar distintos modelos

# Incremento temporal en años (fijo)
incrementoTemporal = 1.0e7  # [años]

# ----------------------------------------------------------------------
# Simulación hacia el pasado (t < 0)
# ----------------------------------------------------------------------

resultadoPasado = []

tiempoActual = 0.0             # [años]
factorEscalaActual = 1.0        # a(t0) = 1
hubbleActual = constanteHubble   # aquí jugamos con ȧ(t0) ≃ H0 (porque a0 = 1)

# Integramos hacia el pasado hasta a ~ 0.01
while factorEscalaActual > 0.01:

    # Aceleración d2a/dt2 según la ecuación de Friedmann (22)
    aceleracion = -(4.0/3.0) * math.pi * constanteGravedad * (
        densidadMateria   / factorEscalaActual**2 +
        2.0 * densidadRadiacion / factorEscalaActual**3 +
        (1.0 + 3.0 * omega) * densidadEnergia / factorEscalaActual**(3.0 * omega + 2.0)
```

```
    )

    # Euler explícito hacia el pasado (dt < 0)
    hubbleActual       = hubbleActual - incrementoTemporal * aceleracion
    factorEscalaActual = factorEscalaActual - incrementoTemporal * hubbleActual
    tiempoActual      -= incrementoTemporal

    # Guardamos tiempo (en Gyr) y factor de escala
    resultadoPasado.append((tiempoActual / 1.0e9, factorEscalaActual))

# ----------------------------------------------------------------------
# Simulación hacia el futuro (t > 0)
# ----------------------------------------------------------------------

resultadoFuturo = []

tiempoActual = 0.0
factorEscalaActual = 1.0
hubbleActual = constanteHubble

# Integramos hasta 10 Gyr hacia el futuro
while tiempoActual < 1.0e10:

    # Aceleración d2a/dt2 (misma expresión que arriba)
    aceleracion = -(4.0/3.0) * math.pi * constanteGravedad * (
        densidadMateria   / factorEscalaActual**2 +
        2.0 * densidadRadiacion / factorEscalaActual**3 +
        (1.0 + 3.0 * omega) * densidadEnergia / factorEscalaActual**(3.0 * omega + 2.0)
    )

    # Euler explícito hacia el futuro (dt > 0)
    hubbleActual       = hubbleActual + incrementoTemporal * aceleracion
    factorEscalaActual = factorEscalaActual + incrementoTemporal * hubbleActual
    tiempoActual      += incrementoTemporal

    resultadoFuturo.append((tiempoActual / 1.0e9, factorEscalaActual))

# ----------------------------------------------------------------------
# Unión de resultados y representación gráfica
# ----------------------------------------------------------------------

# Ordenamos el pasado de más antiguo a t = 0
resultadoPasado.reverse()

# Unimos pasado, presente (a=1 en t=0) y futuro
resultados = resultadoPasado + [(0.0, 1.0)] + resultadoFuturo

data = np.array(resultados)
x, y = data.T

fig, ax = plt.subplots()
ax.plot(x, y)

ax.set_title('Expansión del universo')
ax.set_xlabel('tiempo (relativo al presente, en miles de millones de años)')
ax.set_ylabel('factor de escala (a(t) / a_0)')
ax.legend([f'$\\omega$ = {omega}'], loc='best')
```

```
plt.show()
```